\documentclass[aps,pra,superscriptaddress,floatfix,twocolumn]{revtex4-1}
\usepackage{graphics}
\usepackage[utf8]{inputenc}
\usepackage{amsmath}
\usepackage[demo]{graphicx}
\usepackage{caption}
\usepackage{subcaption}
\usepackage{graphicx}
\usepackage{float}
\usepackage{geometry}
\usepackage{verbatim}
\usepackage{xcolor}

\begin{document}
	
\title{Quantum Liquid in Lower Dimensions: From the perspective of Surface Tension}
\author{Sri Satwika Adusumalli}\thanks{contributed equally}
\affiliation{Department of Physics, IIT Roorkee, Roorkee, 247667, Uttarakhand, India}
\author{Kathaa Senapati}\thanks{contributed equally}
\affiliation{Dipartimento di Fisica e Astronomia ``Galileo Galilei", Universit\`a degli Studi di Padova, Via Marzolo, 8 - 35131 Padova, Italy}
\author{Shivam Singh}\author{Ayan Khan}\thanks{ayan.khan@bennett.edu.in}
\affiliation{Department of Physics, School of Engineering and Applied Sciences, Bennett
University, Greater Noida, UP-201310, India}
\begin{abstract}
We analyze the surface tension in ultra-cold atomic gases in a quasi one-dimensional and one-dimensional geometry. In recent years, experimental observations have confirmed the ``clustering of atoms" to form droplets in ultra-cold atomic gases and the emergence of this new phase is attributed to the beyond mean-field interaction. However, two decades earlier, liquid formation was predicted due to the competition of two-body and three-body interactions. Here, we review both propositions and comment on the role of beyond mean-field and three-body interaction in liquid formation by calculating the surface tension.
\end{abstract}

\maketitle
\section{Introduction}
The physics at ultra-low temperatures is always a matter of extreme interest. In this direction a new impetus has been provided through the experimental observations of liquid-like phase in ultra-cold gases \cite{barbut2,barbut1,chomaz,chomaz1,cabrera1,semeghini,cabrera2}. The observations of this interesting phase were noted in dipolar Bose Eistein condensate (BEC) \cite{barbut2,barbut1,chomaz,chomaz1} and in binary BECs \cite{cabrera1,semeghini,cabrera2}. These experimental observations were treated as verification of the suggested theoretical framework offered recently \cite{petrov1,petrov2}.
In the current theoretical model, the beyond mean-field (BMF) contribution (proposed by Lee-Huan-Yang \cite{lee}) is considered as the catalyst in the stabilization mechanism which in effect supports the formation of the liquid-like droplets (can be called dropleton). One of the prominent signatures of the formation of liquid-like states is the homogeneous spatial density \cite{barbut3}. 

However, one must note that, about two decades ago, a similar state of matter was predicted based on the Efimov effect which results in the creation of three-body bound states \cite{bulgac1,bulgac2,bulgac3}. The competition between the two-body mean-field interaction (cubic non-linearity) and three-body interaction (quintic non-linearity) is at the heart of this theory. The liquid formation was substantiated through the calculation of non-zero surface tension. Here, it is important to note that in Ref.~\cite{barbut1} the role of three-body interaction is discussed and the authors have concluded that the BMF interaction after dipolar correction stabilizes the system at relatively lower density compared to the three-body contribution.

In recent years there have been some suggestions that the flat-top density distribution arising from the competition between cubic-quintic non-linearity, actually points to flat-top solitons (FTS) or platicons which are different from the droplets in its dynamical nature \cite{lobanov,khawaja4}. In this context, we plan to investigate the ambiguity associated with dropleton and platicon by analyzing the non-linear Schrodinger equation with LHY correction and three-body interaction (as suggested in Ref.\cite{bulgac1}) in one-dimensional (1D) and quasi 1D (Q1D) systems. 

The theoretical discussions on droplet formation in 1D Bose gas or Q1D binary BECs revolve around the nature of the BMF interaction. To be more precise, in the 1D system the BMF interaction manifests via quadratic non-linearity \cite{petrov2,malomed1} while it is quartic non-linearity in Q1D \cite{debnath1}. Our primary objective is to study the competition between two-body mean-field interaction, BMF and three-body interaction in liquid formation. For this purpose, we will rely on the calculation of surface tension. 

We have arranged our results in the following way: in Sec.~\ref{theory} we summarize the recent theoretical model to describe Q1D and 1D systems after taking into account the LHY contribution. Later, we note down the equation of motion in Q1D and 1D with three-body interaction, which we describe as cubic-quartic-quintic non-linear Schr\"odinger equation (CQQNLSE) and quadratic-cubic-quintic non-linear Schr\"odinger equaiton (QCQNLSE) respectively. In Sec.\ref{solution} we explicate the analytical solutions and calculate the surface tension. Our findings relate to both dimensions and we identify the role of interaction as well as dimensionality in liquid formation. We draw our conclusion in Sec.\ref{conclusion}.     
\section{Theoretical Models}\label{theory}
In this section, we plan to summarize the existing theories for both Q1D and 1D. The theory of Q1D is derived from the three-dimensional model \cite{petrov1} with a suitable dimension reduction technique (which we will touch upon here) \cite{debnath1} and the 1D theory is provided in Ref.~\cite{petrov2}.
\begin{figure}[H]
\includegraphics[scale=0.25]{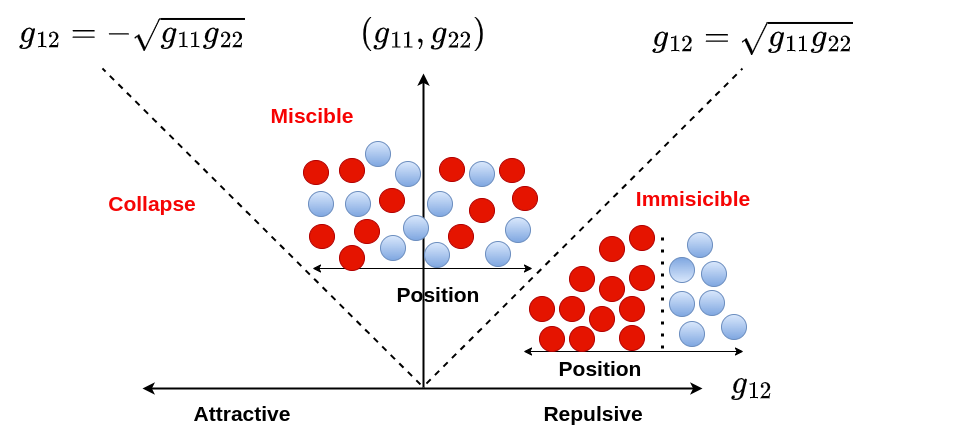}
\caption{(Color Online) Schematic representation of the phase separation due to the competition between inter-species and intra-species interactions in atomic two-component BECs. The red (dark grey in b/w) and blue (light grey in b/w) circles represent two different atomic species.}\label{2bec_schematic}
\end{figure}
In two-component BECs or binary BECs (the two components can be two different species of atoms \cite{pu} or two different hyper-fine states of the same species \cite{myatt}) due to the competition between inter-species and intra-species interactions one can observe distinct mixing dynamics (as schematically described in Fig.~\ref{2bec_schematic}). The intra-species coupling is described via $g_{11}$ and $g_{22}$ while the inter-species interaction is denoted as $g_{12}$. If $g_{11}, g_{22}$ and $g_{12}$ all are repulsive then we can look forward to the transition between the miscible and the immiscible phase. The condensate will be in miscible phase when $- \sqrt{g_{11} g_{22}} <  g_{12}  <  \sqrt{g_{11}g _{22 }}$, and will be in immiscible phase when $g_{12} > \sqrt{g_{11} g_{22}}$ while the condensate will collapse for $ -\sqrt{g_{11} g_{22}} > g_{12}$.

In a uniform Bose mixture, having two components of masses $m_{1}$ and $m_{2}$ \cite{martipi}, 
the coupling constant of intra-species and inter-species interactions are defined as $g_{11} = 4 \pi a_{11} \hbar^{2}/m_{1},\hspace*{2mm} g_{22} = 4 \pi a_{22} \hbar^{2}/m_{2} $ and $g_{12} = 2 \pi a_{12} \hbar^{2}/ m_{r} $ respectively. Here, $m_{r}$ is the reduced mass ($m_{1} m_{2} / (m_{1} + m_{2}$) and, $a_{11} $ and $a_{22}$ are the intra-species $s$-wave scattering length. It is assumed that both these scattering lengths are positive, and the inter-species scattering length ($a_{12}$) is negative. If number densities are $n_{1}$ and $n_{2}$ then their normalized form can be represented as : $\int_{V} n_{1} dr = N_{1}$ and  $\int_{V} n_{2} dr = N_{2} $ respectively. Here, $N = N_{1} + N_{2}$ represents the total number of bosons in the mixture along with the total number density, $ n = n_{1} + n_{2} $. Therefore, energy functional can be written as \cite{salasnich3},
\begin{eqnarray}
E&=&\sum_{j = 1,2}^{} \Big[ \frac{\hbar^{2}}{2m_j}|\nabla \psi_{j}|^{2}+ V_{ext}(r)|\psi_{j}|^{2} + \frac{1}{2} g_{jj} |\psi_{j}|^{4}  \Big]\nonumber\\
&& + g_{12} |\psi_{1}|^{2} |\psi_{2}|^{2} + E_{BMF} (\psi_{1},\psi_{2}).\label{2bec_efunc}
\end{eqnarray}
Here, $E_{BMF}$ stands for the beyond mean-field interaction energy. As described in Fig.~\ref{2bec_schematic}, near the line of collapse where the effective interaction $\delta g=g_{12}+\sqrt{g_{11}g_{22}}\rightarrow 0$, the relatively weak BMF interaction plays important role in stabilizing the system. For the case when we had gone close to the MF instability boundary i.e $\delta g < 0 $, we need to avoid the imaginary contribution for which we approximate that the instability is very weak i.e $|\delta g| \ll g$ and hence we assume $|\delta g| \approx 0$.

Now, considering that density ratio will be $n_{1}/n_{2} = \sqrt{g_{22}/g_{11}}$ = $\eta $, via two components have indistinguishable spatial mode $n_{j=1,2} = |\psi|^{2} $. The energy functional in Eq.(\ref{2bec_efunc}) can now be rewritten as (assuming $m_1=m_2=m$),
\begin{eqnarray}
E&=&\frac{\hbar^{2}}{2m}|\nabla \psi|^{2} + V_{ext}  |\psi|^{2} + \delta g \frac{\eta}{(\eta + 1)^{2}}  |\psi|^{4}\nonumber\\&& + \frac{8 m^{3/2}}{15 \pi^{2} \hbar^{3}}( \eta g_{11} )^{5/2} |\psi|^{5}
\end{eqnarray}
The corresponding effective one-component equation of motion can be noted as \cite{cabrera2}, 
\begin{equation}
i \hbar \frac{\partial \psi }{\partial t } = \Big[ - \frac{\hbar^{2}}{2m} \nabla^{2} + V_{ext} + U |\psi|^{2} + U^{'} |\psi|^{3} \Big] \psi,\label{eom_1c}
\end{equation}
where, $U$ = $ 2 \delta g \frac{\eta}{(1+\eta)^{2}} n$ and $U^{'} $ = $\frac{4 m^{3/2}}{3 \pi^{2} \hbar^{3}} (n g_{11})^{5/2} n^{3/2} $. Here, it is important to note that the above equation contains two types of non-linearity: one is cubic non-linearity, which is responsible for the mean field (MF) interaction and another is quartic non-linearity which relates to beyond mean field (BMF) interaction.

\subsection*{Quasi-One-Dimensional System}
In Eq.(\ref{eom_1c}), the trap potential $V_{ext}$ has two components: one in the transverse direction, defined as $V_{T}(y,z) = \frac{1}{2} m \omega^{2}_{\perp} (y^{2} + z^{2})$ and the other in the longitudinal direction, noted as $V_{L}(x) = \frac{1}{2} m \omega_{o}^{2} x^{2}$ where $\omega_{\perp}$ and $\omega_{o}$ is the transverse and longitudinal trap frequency respectively \cite{khaykovich}.
One can now conveniently reduce Eq.(\ref{eom_1c}) from $3+1$ dimension to $1+1$ dimension by setting $\omega_{\perp}$ as ten times the $\omega_{o}$. Hence, by tuning the trapping frequency it is possible to create a cigar-shaped BEC in a quasi-one dimension.
For a trapped gas the characteristic length scale is defined as $a_{\perp} = \sqrt{\frac{\hbar}{m \omega_{\perp}}}$, however, in quasi one dimension it would be, $a_{x}/a_{\perp} \approx \sqrt{10}$ \cite{debnath1}. This will imply that the transverse trapping frequency is stronger than the longitudinal trapping frequency allowing the condensate to expand in one direction. Here, $a_{x} $ will be $ a_{x} = \sqrt{\frac{\hbar}{m \omega_{o}}}$. In order to reduce the dimension of Eq.(\ref{eom_1c}) in an effective one dimensional (Q1D) equation we consider an ansatz of the following form \cite{khan1},
\begin{equation}
	\psi(r,t) = \frac{1}{\sqrt{2 \pi a_{B} a_{\perp}}} \psi \Big( \frac{x}{a_{\perp}}, \omega_{\perp} t\Big) e^{\Big(-i \omega_{\perp} t - \frac{y^{2} + z^{2}}{a_{\perp}^{2}}\Big)}
\end{equation}

A systematic and careful calculation then leads to the Q1D CQNLSE (here, we have ignored the details for brevity),
\begin{eqnarray}
i \frac{\partial \psi(r,t)}{\partial t }&=&\left[- \frac{1}{2} \frac{\partial^{2}\psi(x,t)}{\partial x^{2}} + \frac{1}{2} K x^{2} 
	+ G_{1} |\psi(x,t)|^{2}\right.\nonumber\\&&\left. + G_{4}|\psi(x,t)|^{3}	\right] \psi(x,t),
\end{eqnarray}
where $G_{1} = \frac{\delta g nm }{4 \pi a_{B} \hbar^{2}}$, $G_{4} = \frac{\sqrt{2}}{3} \frac{m^{3/2} n^{3/2} g_{sym}^{5/2}}{\pi^{2} a_{B}^{3/2} a_{\perp} \hbar^{5}}$, $g_{sym}=g_{11}=g_{22}$ and $K = {\omega_{o}^{2}}/{\omega_{\perp}^{2}}$.
In a more generic manner we can write the equation as, 
\begin{eqnarray}\label{cqnlse}
i \frac{\partial \psi(r,t)}{\partial t }&=&\left[- \frac{1}{2} \frac{\partial^{2}\psi(x,t)}{\partial x^{2}} + V_{ext}(x)
	+ g_{1} |\psi(x,t)|^{2}\right.\nonumber\\&&\left. + g_{4}|\psi(x,t)|^{3}	\right] \psi(x,t),
\end{eqnarray}
Here, $g_1$ and $g_4$ stands for the MF and BMF interaction strength while $V_{ext}(x)$ is the external trap potential. 

\subsection*{One Dimensional System}
Contrary to the Q1D picture (derived from a three-dimensional setup), in 1D Bose gas the BMF contribution manifests as quadratic non-linearity. Further, one can also note that the nature of MF and BMF interaction is reversed in 1D, i.e., MF interaction is repulsive and BMF interaction is attractive. 

In a 1D system, we define, $\delta g$ = $g_{12} + \sqrt{g_{11}g_{22}}$ which is in the vicinity of the unstable region which implies $ 0 < \delta g < \sqrt{g_{11} g_{22}}$. 
Therefore, to explain the one-dimensional GP equation including the beyond mean field (BMF) correction, a small variation between the inter-component attraction strength $(g_{12} < 0) $ and intra-component self-repulsion, $\delta g = g + g_{12} > 0 $ along with $g < 0 $, can be expressed as \cite{debnath6},
\begin{eqnarray}\label{eq: One_dimension}
	i \hbar \frac{\partial \psi}{\partial t}&=&- \frac{\hbar^{2}}{2m} \frac{\partial^{2} \psi}{\partial x^{2}} + V_{ext}(x)\psi + \delta g |\psi|^{2} \psi \nonumber\\&&- \frac{\sqrt{2m}}{\pi \hbar} g^{3/2} |\psi| \psi,
\end{eqnarray} 
where $V_{ext}(x)$ is the trap potential and $m$ is the atomic mass. For further analysis, we write the equation motion in a more generic form such that, 
\begin{equation}\label{qcnlse}
	i \frac{\partial \psi}{\partial t} = - \frac{1}{2}   \frac{\partial^{2} \psi}{\partial x^{2}} + V_{ext}(x)\psi + g_3|\psi|^{2} \psi + g_2|\psi| \psi.
\end{equation}
Here, $g_3$ and $g_2$ denote the strength of the MF and BMF interactions respectively.
\subsection*{Inclusion of Three-Body Interaction}
In the earlier subsections, we have discussed the inclusion of BMF contribution in the GP equation at lower dimensions. However, exactly two decades ago, the first suggestion of droplet formation due to the competition between two-body MF interaction and three-body interaction came up \cite{bulgac1,bulgac2,bulgac3}. Here, we plan to discuss the formalism introduced in Ref.~\cite{bulgac1}.

It is well known that for a dilute Bose system if the two-body scattering length is very large i.e., $a \gg r_{0}$, where $r_0$ is the range of the potential, we start observing three-body bound state formation \cite{bulgac1}. This is commonly described as the Efimov effect \cite{efimov}. In this regime, there are two independent dimensionless parameters $\rho a^3$ and $\rho r_0^3$. The three-body bound
states appear irrespective of whether the two-body scattering length is positive or negative. It is further noted that, if the three-body zero-energy scattering amplitude is defined as $g_3$ then the contribution of the three-particle collisions to
the ground state energy density of a dilute Bose gas can be expressed as, $\mathcal{E}_3(\rho)=\frac{1}{6}g_3\rho^3$. In this situation, 
the ground state energy for N bosons in an ensemble can be noted as,
\begin{equation}\label{eq: Surface tension}
E(N) = \int d^{3}r \Big[\frac{\hbar^{2}}{2m} |\nabla \psi(r)|^{2}  + \frac{1}{2} g_{1} \rho^{2}(r) + \frac{1}{6} g_{3} \rho^{3}(r) \Big]
\end{equation}
Here, $\rho(r) = |\psi(r)|^{2}$ is the number density. Also $g_{1} = {4 \pi \hbar^{2} a}/{m}$ ($ a $  = two particle scattering length and $ m $ = particle mass).

Following the usual prescription of energy minimization and applying the dimension reduction method expressed earlier, we can write the equation of motion as, 
 \begin{equation}\label{eq:CQuinticNLSE}
i \hbar \frac{\partial \psi }{\partial t } =  - \frac{\hbar^{2}}{2m} \frac{\partial^{2} \psi }{\partial x^{2}} + \frac{g_{1}}{2}|\psi|^{2} \psi + \frac{g_{3}}{6} |\psi|^{4} \psi
 \end{equation}

However, accounting the result from Eq.(\ref{cqnlse}) and Eq.(\ref{qcnlse}) and incorporating the three-body interaction in these equations the effective cubic-quartic-quintic NLSE (CQQNLSE) and quadratin-cubic-quintic NLSE (QCQNLSE) will read: 
\begin{eqnarray}
i \frac{\partial \psi }{\partial t} &=& - \frac{1}{2} \frac{\partial^{2} \psi }{\partial x^{2}} + g_{1} | \psi |^{2}\psi + g_{4} |\psi|^{3} \psi \nonumber\\
&&+ g_{3} |\psi|^{4} \psi + V_{ext}(x) \psi\label{cqqnlse}\\
i \frac{\partial \psi }{\partial t} &=& - \frac{1}{2} \frac{\partial^{2} \psi }{\partial x^{2}} + g_{1} | \psi |^{2}\psi + g_{2} |\psi| \psi \nonumber\\
&&+ g_{3} |\psi|^{4} \psi + V_{ext}(x) \psi\label{qcqnlse}
\end{eqnarray}   
where, $g_{1}$ and $g_{3}$ are the two-body and three-body interaction strength respectively while, $g_{2}(g_4)$ describes the BMF interaction in 1D (Q1D) geometry. Further, the trap potential is denoted as $V_{ext}(x)$.
In the next section, we will propose a possible analytical solution. Further, we will comment on the potential structures which are essential to stabilize the solutions. 
Also in the above equations, we have not made any specific comment on the nature of the interactions which we plan to explicate in the next section.

\section{Analytical solutions}\label{solution}
In the preceding section, we have discussed the equation of motion in different geometries and explained the intriguing nature of the BMF interaction while moving from a Q1D to a 1D system. One can find a more detailed discussion on this aspect in Ref.~\cite{debnath6}. Here, we plan to examine the existence of cnoidal solutions for the above-mentioned equations. For this purpose, we first discuss Q1D geometry and then we elaborate on the 1D system.
\subsection*{Analytical Solution in Q1D}  
Now, we are interested in finding out the solution of Eq.(\ref{cqqnlse}) by considering trapping potential in a cnoidal form \cite{abra,debnath6}. Further, we assume the temporal evolution is sinusoidal such that, $\psi (x,t) = \phi(x) e^{-i \mu t}$. Here, $\mu$ can be interpreted as chemical potential. We also use $z = \xi x$ so that $z$ is dimensionless and $\xi$ can be interpreted as the inverse of the coherence length. The cnoidal potential is defined as, $V(z)=V_{0} cn^{4}(z,m)$, where $V_{0}$ is strength of the trap potential.  The form of the potential plays an important role in stabilizing the solution contrary to competing nonlinearities. Hence, the modified CQQNLSE will read,
\begin{eqnarray}
&&\xi^{2} \frac{\partial^{2} \phi}{\partial x^{2}} - \phi(z)[\mu - g_{1} |\phi(z)|^{2} - g_{4} |\phi(z)|^{3} \nonumber\\&&- g_{3} |\phi(z)|^{4} - V_{0} cn^{4}(z,m)] = 0.\label{cqqnlse1}	
\end{eqnarray} 
To find an analytical solution for Eq.(\ref{cqqnlse1}) we consider an ansatz of the form $\phi(z) = A+ B cn(z,m)$, $m$ being the moduli parameter.

So, by applying the ansatz in Eq.(\ref{cqqnlse1}), we derive the set of consistency conditions such that,
\begin{eqnarray}
&&-A \mu + A^{3}g_{1} + A^{4} g_{4} + A^{5} g_{3}=0 \label{eq: Q1D_1}\\
&&-B(\mu + \xi^{2} - 2m\xi^{2} - 3A^{2}g_{1} - 4A^{3}g_{4} - 5A^{4}g_{3} )=0 \nonumber\\&&\label{eq: Q1D_2}\\
&&AB^{2}(3 g_{1}+2A(3g_{4}+5 Ag_{3}))=0 \label{eq: Q1D_3}\\
&&-2Bm\xi^{2}+B^{3}g_{1}+4AB^{3}g_{4}+10A^{2}B^{3}g_{3}=0 \label{eq: Q1D_4}\\
&&B^{4}g_{4} + A(5B^{4}g_{3}+V_{0})=0 \label{eq: Q1D_5}	\\
&&B(B^{4}g_{3}+V_{0})=0 \label{eq: Q1D_6}
\end{eqnarray}
Solving and rearranging the set of equations Eq.(\ref{eq: Q1D_1}) to Eq.(\ref{eq: Q1D_6}), we are able to find solution parameter ($A$ and $B$) in terms of the equation parameters ($V_0$, $g_3$ and $g_4$).
\begin{equation}
	{B = \Big(-\frac{V_{0}}{g_{3}} \Big)^{\frac{1}{4}}}, \hspace{5mm} {A = - \frac{g_{4}}{4g_{3}}}. \label{A&B}
\end{equation}
Further, we obtain a set of constrained condition such as, 
\begin{eqnarray}
g_{1}&=&\frac{7g_{4}^{2}}{24g_{3}}\label{g1}\\
\mu&=&\frac{5 g_{4}^{2}}{768 g_{3}^{3}}\label{chempot_q1d}\\
\xi&=&\sqrt{\frac{g_{4}^{2}}{12g_{3}} \Big[ \frac{g_{4}^{2}}{16g_{3}} - i \sqrt{\frac{V_{0}}{g_{3}}}   \Big]} \label{coherence_q1d}
\end{eqnarray} 
The constrained conditions allow us to examine the competition between the three interaction strengths. From Eq.(\ref{A&B}) and Eq.(\ref{coherence_q1d}) we realize that 
the three body interaction strength ($g_3$) or the strength of the trap potential ($V_0$) must be negative such that the $\xi$ and $B$ are real-valued. This will also ensure that the inverse of coherence length is real. 
Moreover, later we will see that the same condition is also mandatory to obtain real value of surface tension. From the above set of equations, we can also conclude that the nature of BMF interaction 
is redundant in the current context which implies that it can either be attractive or repulsive without making any significant change in the obtained results. Further, for negative $g_3$, from Eq.(\ref{g1}) implies that 
the mean-field interaction is attractive. Similarly, Eq.(\ref{chempot_q1d}) tells us that $\mu$, which can be treated as scaled chemical potential, is negative as well.

\begin{figure}[H]
		\includegraphics[width=0.5\textwidth]{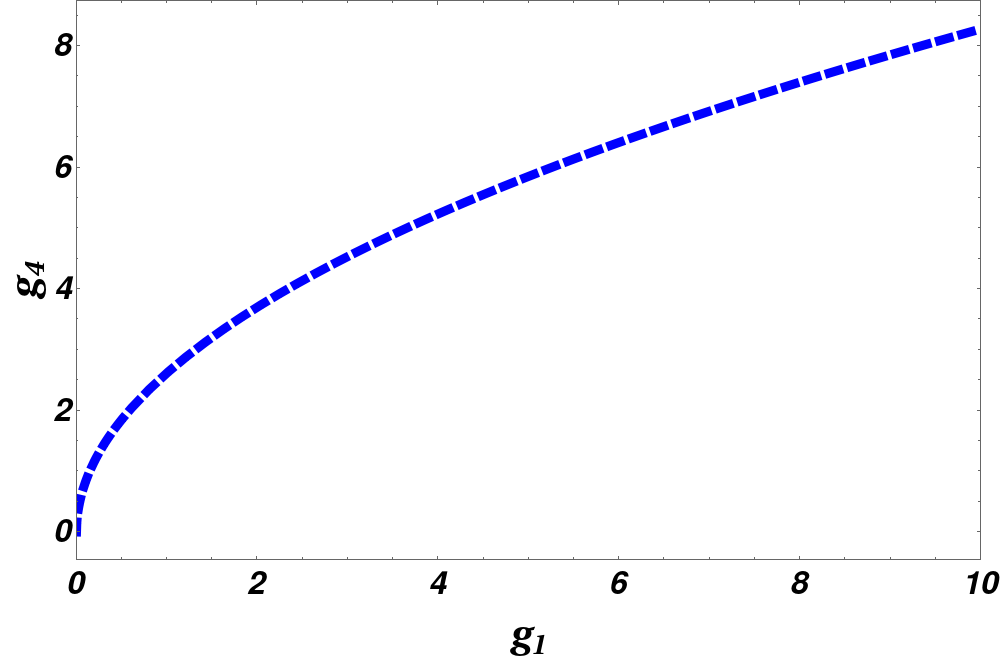}
		\caption{(Color Online) Illustration of parabolic variation of BMF interaction as a function of $g_1$ (for $g_{3}=2$) following Eq.(\ref{g1}). The scale and the parameter value are in an arbitrary unit.}\label{g1_g4}
\end{figure}

\begin{figure}[H]
		\includegraphics[width=0.5\textwidth]{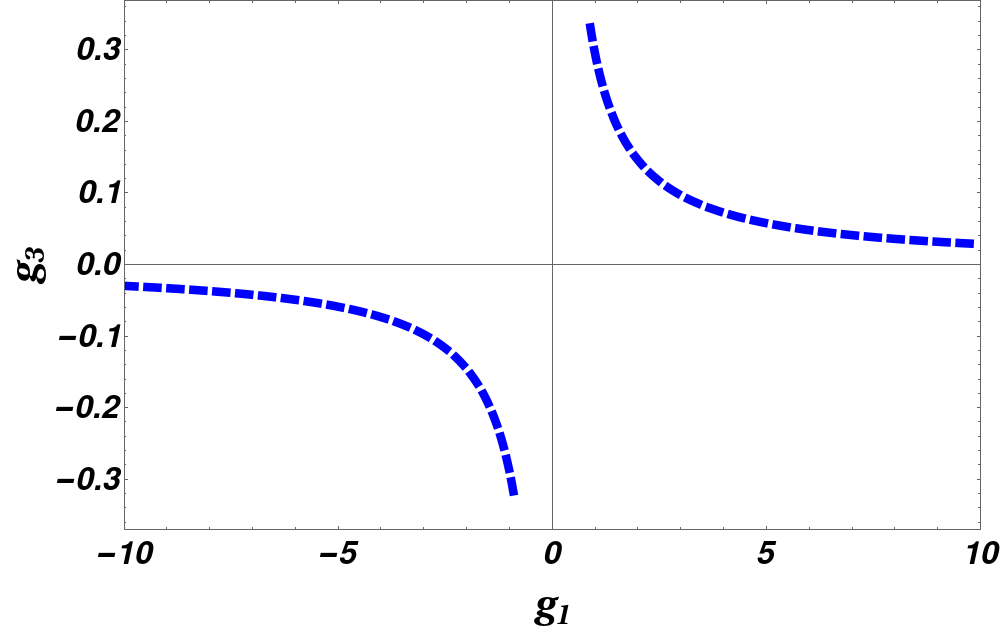}
		\caption{(Color Online) Variation of three-body interaction with two-body interaction strength for a fixed BMF interaction. 
The scale and the parameter values are in arbitrary units. The figure is prepared from Eq.(\ref{g1}) for a fixed BMF interaction ($g_4=1$).}\label{g1_g3}
\end{figure}

From a deeper examination of the interplay between the interactions (as described in Eq.(\ref{g1})), we realize that BMF interaction has parabolic dependence on MF interaction which is depicted in Fig.~\ref{g1_g4}. Interestingly, if BMF interaction is zero then by Eq.(\ref{g1}) the MF interaction is also nonexistent, which leads to infinite coherence length (see Eq.(\ref{coherence_q1d})) and an unacceptable solution. Hence, a small contribution of BMF interaction is essential for a physically meaningful solution. Fig.~\ref{g1_g3} and Eq.(\ref{g1}) reveal that the MF and three-body interactions are inversely proportional which resonates with our intuitions. Also, it is worth noting that $g_4\propto\sqrt{g_1g_3}$.

\begin{figure}[H]
		\includegraphics[width=0.5\textwidth]{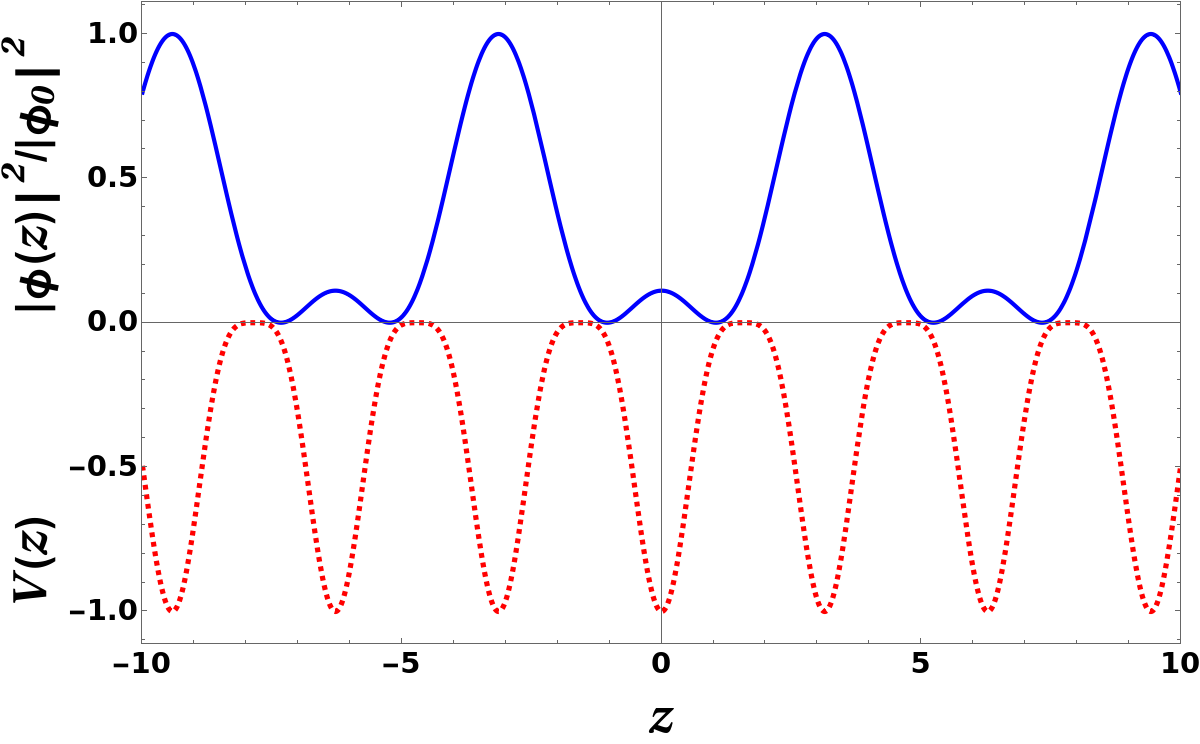}
		\caption{(Color Online) The figure depicts the density distribution (blue solid line) and the variation of the trap potential (red dotted line). The density is scaled by $|\phi_0|^2=|\phi(z=0)|^2$. For better visibility, it is further scaled by the peak value. $V(z)$ is in the units of $V_0$. Here, $m=0$; $g_{4} = 2$; $g_{3} = 1$; $V_{0} =-1$. The parameter values are in arbitrary units.}\label{periodic_q1d}
\end{figure}

\begin{figure}[H]
	\includegraphics[width=0.5\textwidth]{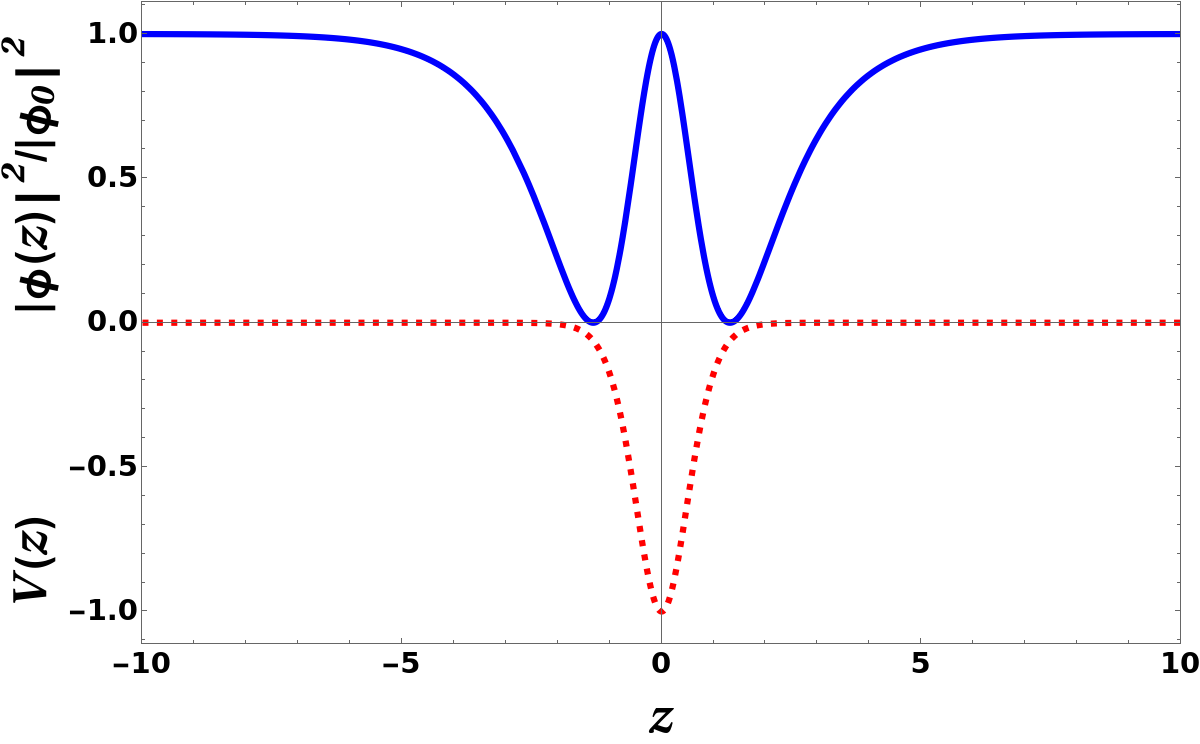}
	\caption{(Color Online) Density distribution (blue solid line) and spatial variation of the trap potential (red dotted line) is described here (scaled by $|\phi_0|^2$). The figure is prepared for $m=1$; $g_{4} = 2$; $g_{3} = 1$; $ V_{0}=-1 $. The parameter values are in arbitrary units.}\label{localized_q1d}
	
\end{figure}

The final solution can be noted as,
\begin{equation}
	{\phi(z) = - \frac{g_{4}}{4 g_{3}} + \Big(-\frac{V_{0}}{g_{3}} \Big)^{\frac{1}{4}} cn(z,m)}.\label{sol_cqqnlse}
\end{equation}
We have depicted the density distribution obtained from Eq.(\ref{sol_cqqnlse}) for $m=0$ and $m=1$ in Fig.~\ref{periodic_q1d} and \ref{localized_q1d} respectively. The density variation is scaled by the density at the centre of trap, i.e., $|\phi_0|^2=|\phi(z=0)|^2=\left(\frac{g_4}{4g_3}\left(\left(\frac{|V_0|}{g_3}\right)^{1/4}\frac{4g_{3}}{g_{4t}}-1\right)\right)^2$.
For $m=0$ we obtain a periodic but binary amplitude distribution. The localized solution corresponding to $m=1$ depicts a $w$-soliton like structure with finite background density \cite{debnath2}.

\subsection*{Analytical Solution in 1D}  
We now repeat a similar strategy to solve QCQNLSE as described in Eq.(\ref{qcqnlse}). We assume, $\psi(x,t) = \phi(z) e^{-i \mu t}$ with ansatz solution as $\phi(z) = B cn(z,m)$.
As described earlier, $z = \xi x$, and $\xi $ is the inverse of coherence length. We realize that to stabilize the system we require a super-lattice type potential such that, 
$V(z) = V_{0} cn(z,m) + V_{1} cn^{4}(z,m)$ \cite{debnath6}. Here, $V_0$ and $V_1$ are the strengths of the two lattice potentials.

Now Eq.(\ref{qcqnlse}) in $z$ variable can be written as,
\begin{eqnarray}
&&\xi^{2} \frac{\partial^{2} \phi }{\partial z^{2}} + \phi(z)[\mu - g_{1}|\phi(z)|^{2}- g_{2}|\phi(z)|-g_{3}|\phi(z)|^{4}\nonumber\\&&-V_{0} cn(z,m) - V_{1} cn^{4}(z,m)] = 0
\end{eqnarray}  

\begin{figure}[H]
		\includegraphics[scale=0.2]{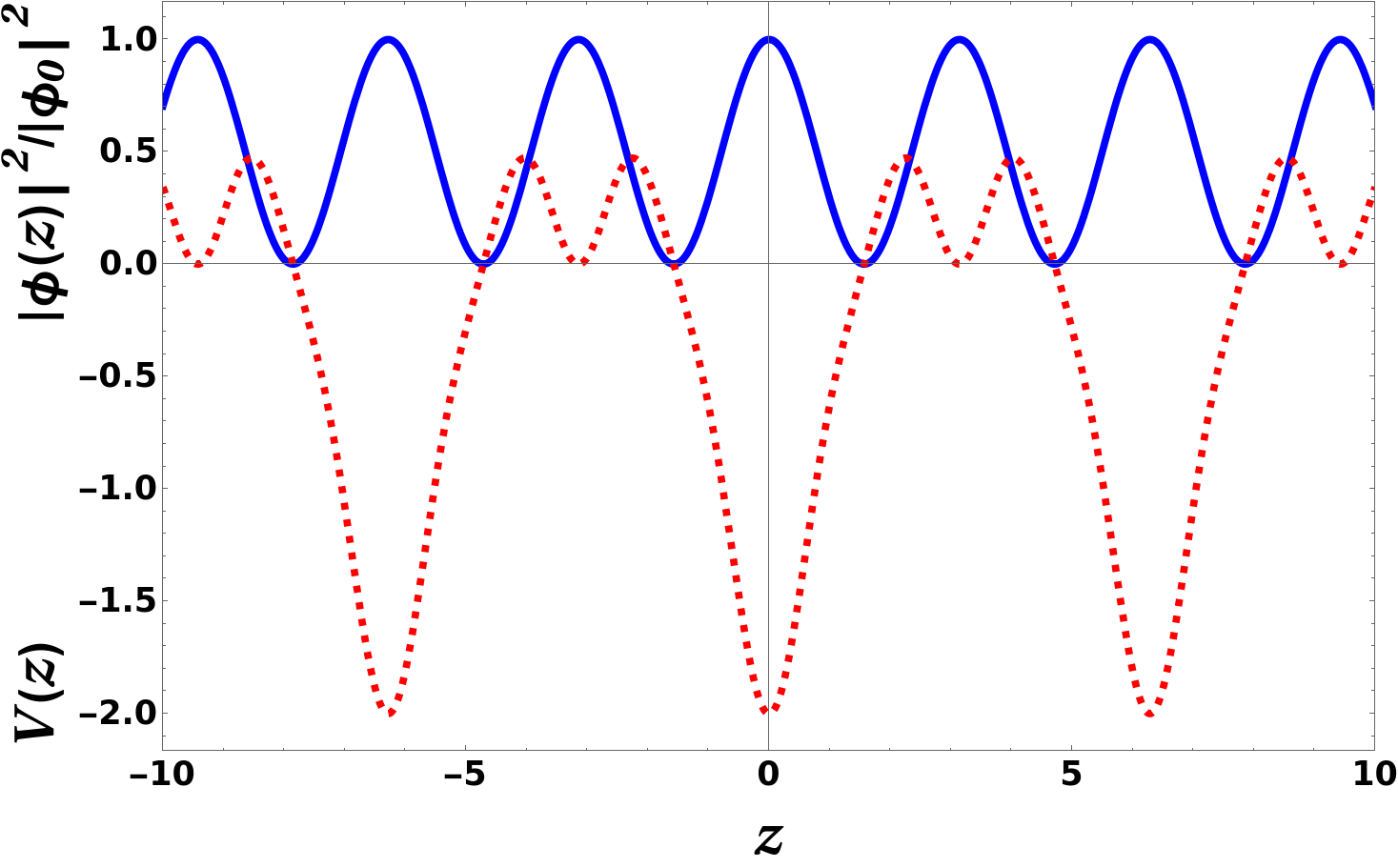}
		\caption{(Color Online) The figure describes the periodic variation of the density mode (blue solid line) in a super-lattice-like potential landscape (red dotted line). The moduli parameter or $m$ is considered as zero here. The density is scaled by the center of the trap density, i.e., $|\phi_0|^2=|\phi(z=0)|^2=(-V_0/g_2)^2$. The other parameters are in arbitrary units such as $g_{2} = 1$; $V_{0} =-1$ and $V_1=1$.}\label{periodic_1d}
\end{figure}

Inserting the ansatz we obtain a set of consistency equations which are noted as,
\begin{eqnarray}
	B \mu - B \xi^{2} + 2 Bm \xi^{2}&=&0\label{eq: 1D_1}\\
	-B^{2}g_{2} - BV_{0}&=&0\label{eq: 1D_2}\\
	-B^{3}g_{1} - 2Bm\xi^{2}&=&0\label{eq: 1D_3}\\
	-B^{5}g_{3} - B V_{1}&=&0\label{eq: 1D_4}
\end{eqnarray}
An appropriate rearrangement will lead to the solution of the Eqs. (\ref{eq: 1D_1}), (\ref{eq: 1D_2}), (\ref{eq: 1D_3}) and (\ref{eq: 1D_4}). 
We yield a relationship between the equation parameters and solution parameters in the following form,
\begin{equation}
	{B = - \frac{V_{0}}{g_{2}}}, \hspace{5mm} {g_{1} = - \frac{2m\xi^{2}}{V_{0}^{2}} g_{2}^{2}}. 
\end{equation} 
Thus, the mean-field interaction is proportional to the square of the BMF interaction. Further, we obtain a constrained relationship between the interactions ($g_2$ and $g_3$) and potential strengths ($V_0$ and $V_1$) such that $\frac{g_{2}^{4}}{g_3}=-\frac{V_{0}^{4}}{V_1}$. Here, it must be noted that the trap potential is independent of the interactions however, the analytical solution necessitates the above-mentioned constrained condition, implying that the numerical value of $\frac{g_{2}^{4}}{g_3}$ should match with $-\frac{V_{0}^{4}}{V_1}$. The chemical potential and the inverse of coherence length reads:
\begin{eqnarray}
\mu&=&- \frac{V_{0}^{2}(1-2m)}{2m} \Big( \frac{g_{1}}{g_{2}^{2}} \Big),\label{chempot_1d}\\
\xi&=&\sqrt{\frac{\mu}{1-2m}}.\label{coherence_1d}
\end{eqnarray} 
\begin{figure}[H]
		\includegraphics[scale=0.2]{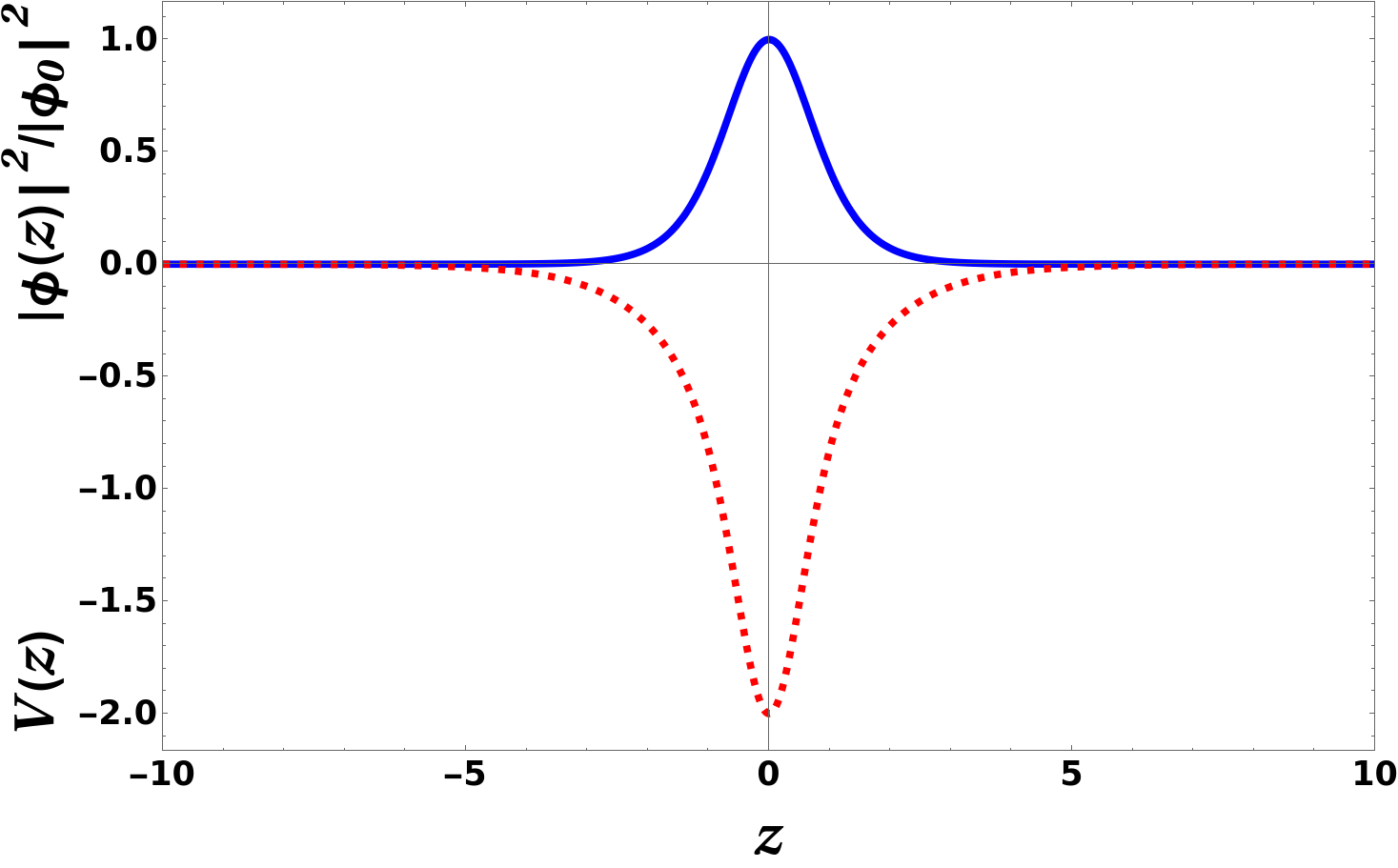}
		\caption{(Color Online) Variation of density (blue solid line) and description of potential landscape (red dotted line) is illustrated here (with moduli parameter, $m=1$). The density is scaled by the center of the trap density, i.e., $|\phi_0|^2=|\phi(z=0)|^2=(-V_0/g_2)^2$. The potential is in units of $V_0$. All the parameters are in arbitrary units such that $g_{2} = 1$; $V_{0} =-1$; $V_1=1$.}\label{localized_1d}
\end{figure}
Therefore, the final solution turns out as,
\begin{equation}
	{\phi(z) = - \frac{V_{0}}{g_{2}} cn(z,m)}\label{sol_qcqnlse}
\end{equation}
A closer inspection of Eqs.(\ref{chempot_1d}) and (\ref{coherence_1d}) allows us to conclude that at $m=0$, $\xi\rightarrow\infty$ so to keep the dimensionless parameter $z$ finite, $x\rightarrow 0$. Also, Eq.(\ref{sol_qcqnlse}) reveals to us that it is essential to have the BMF interaction for a physically acceptable solution. In Fig.~\ref{localized_1d} we depict the density distribution via a blue solid line and the variation of trap potential through a red dotted line. The solution clearly manifests a bright soliton-like nature where $\phi(z)\rightarrow 0$ asymptotically. We also note the periodic solution against the backdrop of the quasi-periodic potential landscape in Fig.~\ref{periodic_1d}.  

\subsection*{Calculation of Surface Tension}
Surface tension can be defined as a tendency of liquid to shrink its surface to a minimum surface area due to cohesive force. Hence, in the context of liquid formation in quantum gases, the existence of surface tension can be regarded as one of the significant pieces of evidence as pointed out in Ref.~\cite{bulgac1}.

In quantum gases, if the effective scattering length is attractive (say $a<0$) then energy per particle relation for a uniform condensate(i.e $E_{2} = \frac{1}{2} g_{2} \rho $, where $g_{2} = \frac{4 \pi \hbar^{2} a}{m} $), is unbounded and density be will unstable with respect to the density fluctuations. However, it was suggested that the inclusion of the three-body correlation effect changes the situation dramatically and predicted that stable droplet formation can be possible via surface tension calculation \cite{bulgac1}.

As noted in previous section, if two body scattering length is large enough(i.e $| a| \gg r_{0}$, where, $r_{0} $ is the two body interaction radius) then there will be two independent dimensionless parameter $\rho a^{3}$ and $\rho r_{0}^{3}$ and we observe the emergence of the three-body bound state \cite{efimov}. In this situation, the surface tension can be defined as \cite{bulgac1},
\begin{equation}
	\sigma = \int_{- \infty}^{\infty} dx [\epsilon(x)-\mu_{0} \rho(x)],\label{ST}
\end{equation}
where $\epsilon(x)$ is the energy density, $\rho(x)$ is the number density and $\mu_{0}$ is the chemical potential corresponding to infinite matter at equilibrium density \cite{bulgac1}. Now applying Eq.(\ref{sol_cqqnlse}) and Eq.(\ref{sol_qcqnlse}) in Eq.(\ref{ST}) using the energy functional corresponding to Eq.(\ref{cqqnlse}) and Eq.(\ref{qcqnlse}) we can obtain the non zero surface tension for Q1D and 1D respectively, 
 \begin{eqnarray}
 	\sigma_{Q1D}&=&\frac{1}{3} \sqrt{- \frac{V_{0}}{g_{3}}},\label{st_q1d}\\
\sigma_{1D}&=&\frac{1}{3}\left( \frac{V_{0}}{g_{2}} \right)^{2}.\label{st_1d}
 \end{eqnarray} 
The calculation is carried out after carefully transforming the $x$ variable to the $z$ variable.

\begin{figure}[H]
		\includegraphics[width=0.5\textwidth]{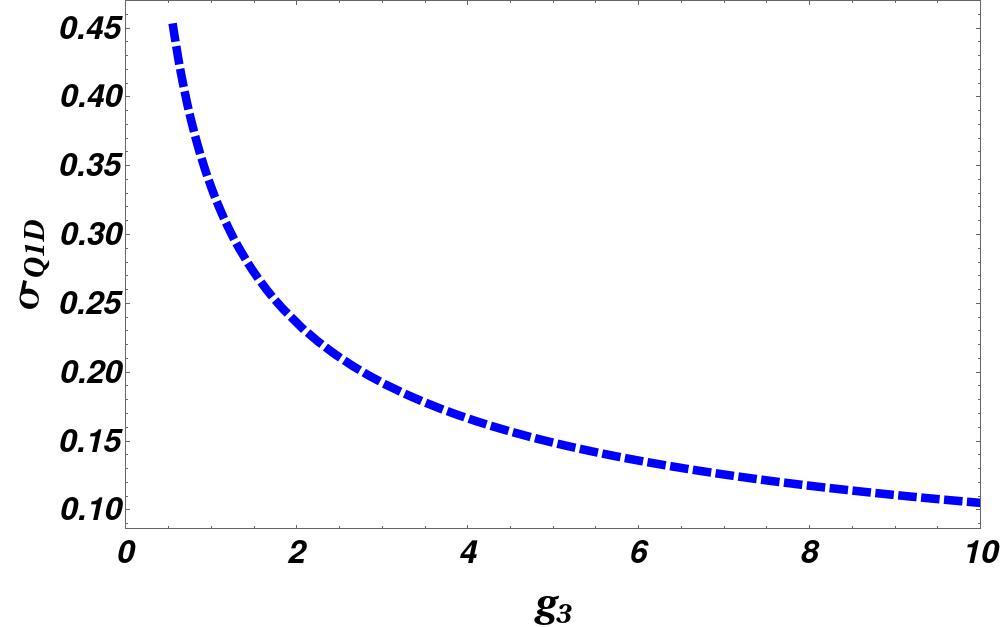}
		\caption{(Color Online) Variation of surface tension as a function of three-body interaction in Q1D system (following Eq.(\ref{st_q1d})). Here, $V_0$ is considered as $-1$ in arbitrary units.}\label{fig_st_q1d}
\end{figure}

\begin{figure}[H]
		\includegraphics[width=0.5\textwidth]{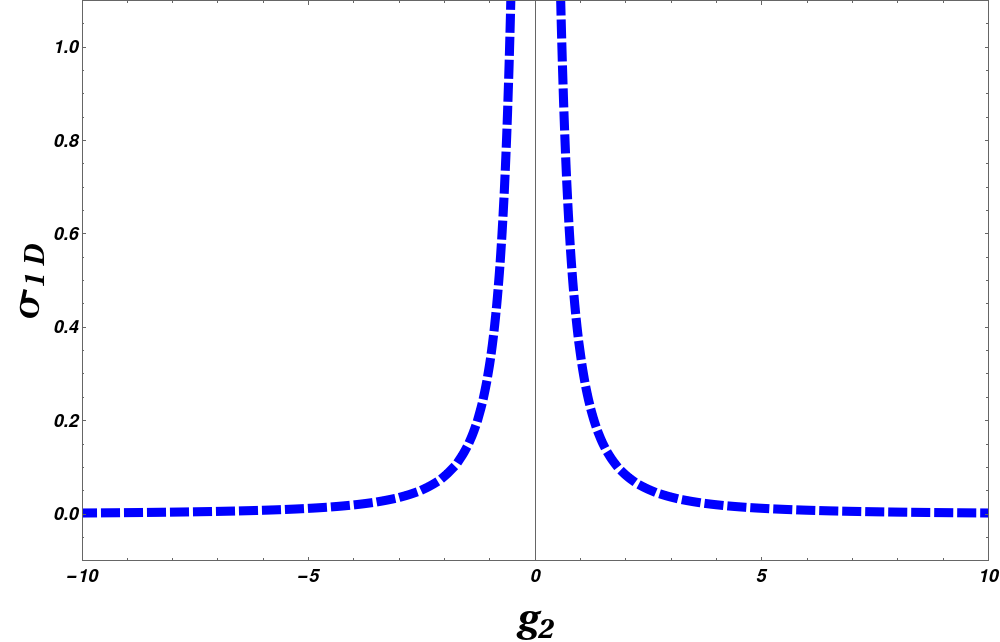}
		\caption{(Color Online) Variation of surface tension with BMF interaction in the 1D system (following Eq.(\ref{st_1d})). Here, $V_0$ is considered as $-1$ in arbitrary units.}\label{fig_st_1d}
\end{figure}

Eq.(\ref{st_q1d}) and Eq.(\ref{st_1d}) possess some common attributes while they are quite distinctive as well. In both cases we find that the trapping potential has an important role to play and if $V_0=0$ then surface tension is zero which implies that the atoms can not form clusters to emerge as liquid-like states. However, one striking difference is, that in Q1D three-body interaction is essential for non-zero surface tension while in 1D the BMF interaction plays that role. Fig.~\ref{fig_st_q1d} describes that the surface tension is proportional to the inverse square root of three-body interaction while in Fig.~\ref{fig_st_1d} we observe that the surface tension varies with the inverse square of BMF interaction. This implies that, in Q1D we should be able to yield quantum droplets even if we do not take into account BMF interaction which was precisely pointed out in Ref.~\cite{bulgac3}, nevertheless in the 1D system, we see a role reversal.

If we now look into the chronological perspective, the recent discussions on droplets are mainly from the perspective of BMF interaction; the earlier discussions were more in the context of the interplay between two-body and three-body interactions \cite{bulgac1,bulgac2}. However, our current discussion suggests that BMF interaction is irreplaceable in 1D, while in Q1D, three-body interaction can induce liquid formation. Hence, based on the recent discussions on platicons (FTS) and dropletons \cite{lobanov,khawaja4}, our current observation, motivates us to characterize the Q1D liquid as platicon and 1D liquid as dropleton.    

\section{Conclusion}\label{conclusion}
In recent years, we have observed significant interest in understanding liquid-like state formation in ultra-cold atomic gases due to the theoretical developments \cite{petrov1,petrov2,petrov3} and experimental verification \cite{cabrera1,barbut1,santos1}. The fundamental theoretical proposition suggests that the BMF interaction competes with the effective MF interaction to support the clustering of atoms. However, two decades ago a similar exotic phase was predicted theoretically by considering the competition of two-body and three-body interactions \cite{bulgac1,bulgac2}. In this article, we have tried to review both the perspectives in Q1D and 1D geometry through a pathological model.

We have already noted interesting differences in Q1D and 1D geometry from mathematical and physical perspectives \cite{debnath6} when we include BMF interaction. Recent investigations reveal that in Q1D, the dynamical equation boils down to a CQNLSE \cite{debnath1} while in 1D it is QCNLSE \cite{petrov2}. We have now included the three-body interaction in the existing Q1D and 1D model and obtained the cnoidal solutions. We know that the cnoidal functions can boil down to periodic or localized functions based on their moduli parameters. Hence, to exploit this mathematical manoeuvrability we provide our solutions to interms of cnoidal functions only. Later, we calculate the corresponding surface tension associated with each dimension. Interestingly, we note that a trap potential is invariably required for non-zero surface tension in both cases. However, in Q1D, surface tension is independent of BMF interaction and we can have non-zero surface tension due to the competition between two-body and three-body interactions while in 1D, it is essential to have BMF interaction whereas three-body interaction appears redundant. 
We hope our current analysis will shed some light on the context of dropleton and platicon and will also motivate some experiments to study these exotic phases in lower dimensions.

\section*{Acknowledgement}
AK thanks the Council of Scientific and Industrial Research (CSIR) Human Resource Development Group (HRDG) Extramural Research Division (EMR-II), India for the support provided through project number 03/1500/23/EMR-II.

\bibliographystyle{apsrev4-1}
\bibliography{ms_v1}

\end{document}